# Stimulated Emission Depletion (STED) Magnetic Particle Imaging

Guang Jia[1,*,#], Zhongwei Bian[2,*], Tianshu Li[3], Shi Bai[3,#], Chenxing Hu[1], Lixuan Zhao[4], Peng Gao[4], Tanping Li[4,#], Hui Hui[5], Jie Tian[5,#]

[1]School of Computer Science and Technology, Xidian University, Xi'an, Shaanxi, China;

[2]School of Biological Science and Medical Engineering, Beihang University, Beijing, China;

[3]School of Information Science and Engineering, Shenyang University of Technology, Shenyang, Liaoning, China;

[4]School of Physics, Xidian University, Xi'an, Shaanxi, China;

[5]CAS Key Laboratory of Molecular Imaging, Institute of Automation, Chinese Academy of Sciences, Beijing, China

*These authors equally contributed to this work.

#Corresponding author: Guang Jia (gjia@xidian.edu.cn), Shi Bai (baishi@sut.edu.cn), Tanping Li (tpli@xidian.edu.cn), Jie Tian (jie.tian@ia.ac.cn)

# Title Page

**Article type:** Research article

**Title:** Stimulated Emission Depletion (STED) Magnetic Particle Imaging

**Running title:** STED-MPI

## Abstract

Magnetic particle imaging (MPI) is an in-vivo imaging method to detect magnetic nanoparticles for blood vessel imaging and molecular target imaging. Compared with conventional molecular imaging devices (such as nuclear medicine imaging PET and SPECT), magnetic nanoparticles have longer storage periods than radionuclides without ionizing radiation. MPI has higher detection sensitivity compared with MRI. To accurately locate molecular probes in living organisms, high-resolution images are needed to meet the requirements of precision medicine. The spatial resolution of the latest domestic and international MPI equipment is 1-6 mm and has not yet met the requirements of medical imaging detection. We previously studied the spatial encoding technology based on pulsed square wave stimulation, which significantly improved the image resolution along the field free line (FFL) direction. This study proposes an innovative idea of high-resolution MPI based on stimulated emission depletion (STED) of magnetic nanoparticle signals. The stimulated emission was implemented by using cosine stimulation on FFL-based MPI scanner systems. The STED signal was generated by adding an offset magnetic field parallel to the FFL, which may form a donut-shaped focal spot or a regular Gaussian focal spot depending on the offset field strength. Focal spot modulation techniques and deconvolution algorithms were developed to improve image resolution.

# Introduction

Magnetic particle imaging (MPI) is an in-vivo imaging method to detect magnetic nanoparticles without ionizing radiation for blood vessel imaging and molecular target imaging[1]. Compared with conventional molecular imaging devices (such as nuclear medicine imaging PET[2] and SPECT[3]), magnetic particles have longer storage periods than radionuclides without ionizing radiation[4], and higher detection sensitivity[5] compared with MRI. To accurately locate molecular probes in living organisms, high-resolution images are needed to meet the requirements of precision medicine[6].

Magnetic particle imaging was only proposed at the beginning of this century, with a short development time and great space for improvement. Gleich proposed The concept of magnetic particle imaging was proposed in 2001. The first magnetic particle imaging scanner was successfully developed in 2005[7]. So far, magnetic particle imaging has only been developed for more than two decades[8], and only small animal magnetic particle imaging scanners are commercially available[9]. The image resolution of current magnetic particle imaging devices is only 1-6 mm (Table 1)[10]. The recently developed human-sized MPI scanners allow a spatial resolution of about 5 mm to 10 mm[11], but clinical applications require sub-millimeter resolution. How to improve the resolution of magnetic particle imaging to sub-millimeters is a critical scientific problem and a technical challenge.

| Table 1. Performance of current MPI scanners | | | | | |
|---|---|---|---|---|---|
| MPI Device Performance | Institute of Automation, Chinese Academy of Sciences | The University of California, Berkeley | Bilken University | Hamburg University | Korea Gwangju University of Science and Technology |
| Resolution | 1mm | 2mm | 2.5mm | 6mm | 4~6mm |
| Field of View | 20cm×20cm×100cm | 60cm×120cm | 3.4cm×1.8cm | 10cm×14cm | 14.9cm×13.8cm×9.8cm |
| Gradient Field | 1-4T/m | 3T/m & 6.2T/m | 0.6T/m | 0.2T/m | 2.5T/m |
| Stimulation Field strength | 30mT@10kHz & 500Hz | 15mT@45kHz | 0.9mT@26kHz | 6mT@25.69kHz | 2.1mT@18.4kHz |
| Sensitivity | 121.9nMol | 76.8nMol | 11μg | 263pMol | 20μg |
| Speed | 1f/m | 1f/m | 2f/s | 2f/s | 4f/s |

Many methods have been developed to improve the resolution of magnetic particle imaging (Figure 1). For example, the method of using superconducting coils to improve the magnetic field gradient[8,10], which requires additional complex hardware. Pulsed stimulation was used to overcome the relaxation barrier of large-sized magnetic particles[12]. This method has specific requirements for stimulation coils and it is difficult to produce a standard pulsed waveform. A low stimulation field amplitude was chosen to suppress greatly on signal-to-noise ratio[13]. Superferromagnetic chain particles can improve the resolution by 10 times[14,15], however, the stability and bio-compatibility of such particles need to be improved. Methods to improve the resolution also include calibrated reconstruction algorithms and AI algorithms. For example, combining magnetic field free lines and a deconvolution algorithm can improve the resolution of the reconstructed image[16]. The calibration algorithm of the system matrix is used to improve the resolution of[17], which is usually time-consuming and laborious, and requires a rapid acquisition method[18]. GPU-accelerated parallel image reconstruction algorithms were used to improve the image resolution[19], however, these algorithms are difficult to exceed the physical resolution limit. Many recent machine learning and deep learning algorithms have been used to reconstruct magnetic particle images[20]. Machine learning algorithms such as K-means++ for the resolution improvement in magnetic particle imaging in mouse islet transplantation models[21]. Deep learning algorithms such as CNN[22] and GAN[23] were used to enhance the image resolution. These algorithms usually require a large amount of training data, and they all lack universality, making it difficult to generalize to other magnetic particle imaging devices or magnetic particle type[24].

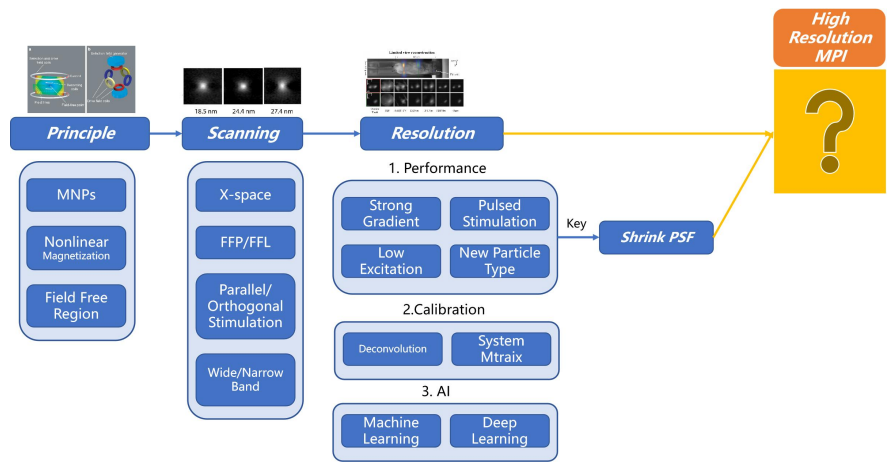

Figure 1. Status of high-resolution magnetic particle imaging methods.

This study put forward innovative research ideas of high-resolution MPI based on stimulated emission depletion (STED) fluorescence microscopic imaging. The theory of STED was proposed by Professor Hell in 1994 to obtain a smaller dot diffusion function (Figure 2), surpassing the traditional optical diffraction limit and achieving a much higher resolution than traditional fluorescence imaging[25]. In 2000, Professor Hell obtained the central dark donut focal spot by using the STED method and achieved a 3-fold resolution improvement[26]. Professor Hell systematically introduced STED fluorescence microscopy based on donut focal spots in 2007, which spans from the "micron" era to the "nano" era[27]. Professor Hell was awarded the Nobel Prize in Chemistry in 2014. Professor Hell's lab introduced the MINSTED technology[28], by changing the radius and position of the donut focal spot, further accurately locating the fluorescent spots and improving the resolution. MINSTED technology was rated as one of the technologies worthy of attention in 2024[29].

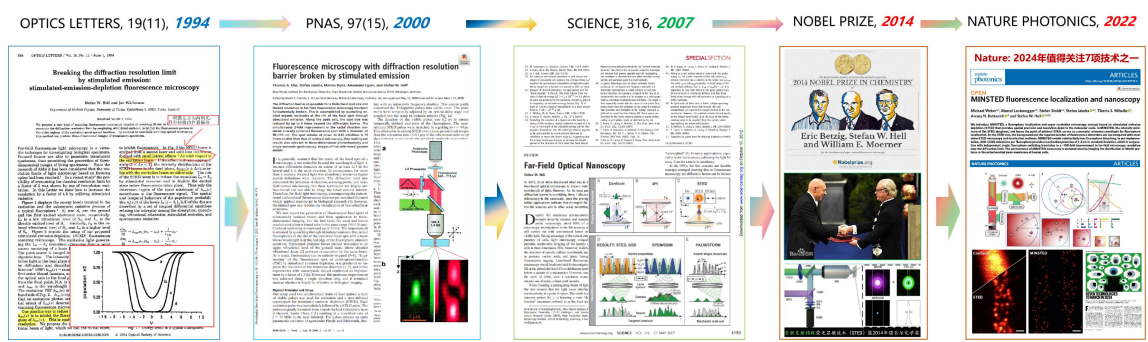

Figure 2. Timeline of STED fluorescence microscopy imaging technology. This study is to use this technology concept to achieve high-resolution magnetic particle imaging.

In this study, we will introduce STED technology to magnetic particle imaging and develop the donut focal spot in magnetic particle imaging (Figure 3). The focal spot shapes under different magnetic particles and stimulation offset conditions will be studied. A focal spot modulation technique will be developed to achieve high-resolution magnetic particle imaging.

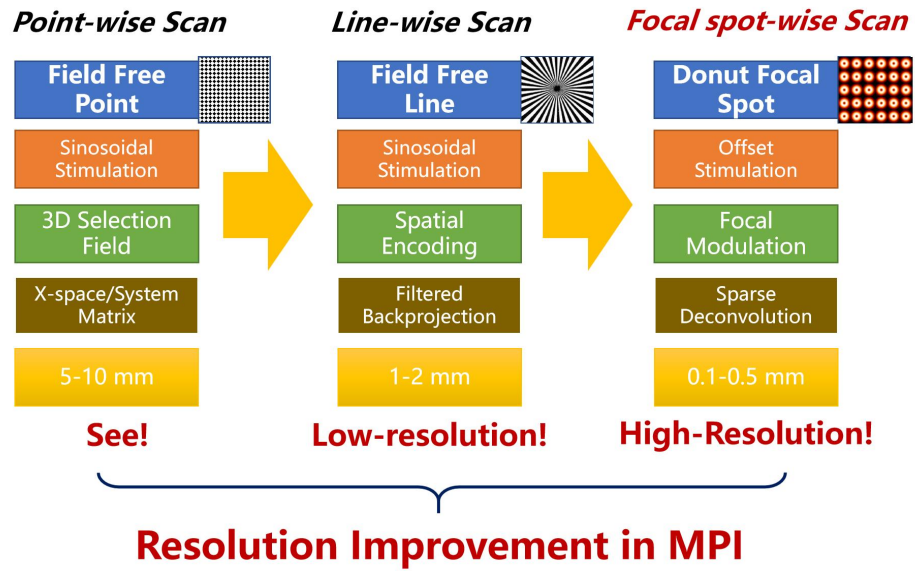

Figure 3. This study is to use the donut-shaped focal spot to improve the image resolution of magnetic particle imaging.

## STED-MPI Theory

*Stimulation Fields*

Magnetic particle spectroscopy (MPS) uses an alternating magnetic field of high field strength to drive the MNP into the nonlinear magnetization response[30,31]. The relaxation mechanisms of magnetization in response to these stimulation magnetic fields can be studied[32]. AC susceptometry (ACS) uses a low time-varying magnetic field with strong offset magnetic fields to investigate the environment of MNPs[33]. Recently, Critical offset magnetic particle spectroscopy (COMPASS) uses a strong stimulation field and a strong offset magnetic field for sensitive and robust investigations of MNP dynamics and surface chemistry[34]. The offset field $H_{DC}$ was parallel to and around one-half of the alternating stimulation field $H_{AC}$. In this study, we combine an excitation field with a parallel strong offset magnetic field and expand the parameter space with an orthogonal gradient field $H_G$ for FFL-MPI (Figure 4).

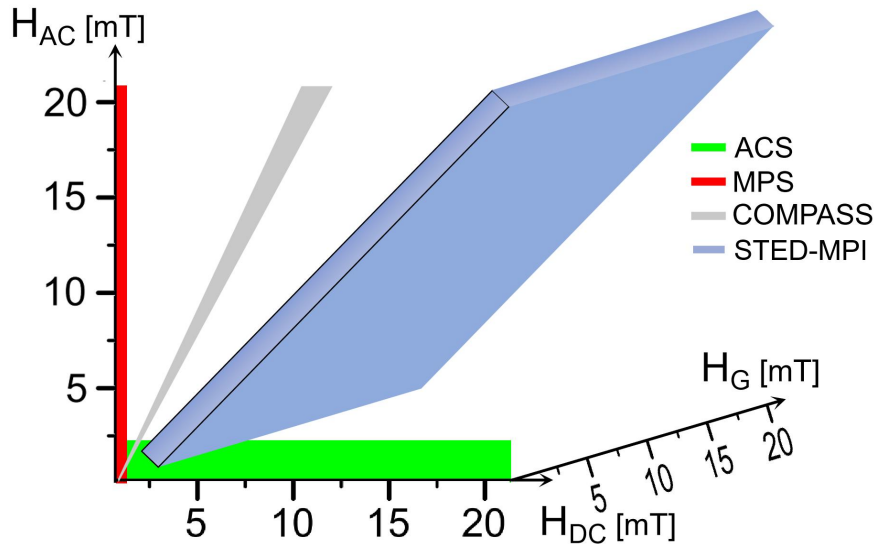

Figure 4. Stimulation parameter space for different MNP measurement and imaging methods.

In an FFL-MPI scanner, the selection gradient field of the FFL along z direction is:

$$H_G(x, y, z) = \begin{pmatrix} G & 0 & 0 \\ 0 & G & 0 \\ 0 & 0 & 0 \end{pmatrix} \cdot \begin{pmatrix} x \\ y \\ z \end{pmatrix} = \begin{pmatrix} Gx \\ Gy \\ 0 \end{pmatrix}, \quad (1)$$

in which G is the magnetic field gradient in both *x* and *y* directions (Figure 5).

If the alternating excitation field has a frequency $\omega_0 = 2\pi/T$ and an amplitude of $H_{AC}$.

$$H_{EX}(t) = \begin{pmatrix} 0 \\ 0 \\ -H_{AC}\cos\omega_0 t \end{pmatrix}. \tag{2}$$

We add an offset field $H_{DC}$ in z-direction,

$$H_{offset} = \begin{pmatrix} 0 \\ 0 \\ H_{DC} \end{pmatrix}. \tag{3}$$

The combined stimulation field is

$$H_{ST}(t) = H_{offset} - H_{EX}(t) = \begin{pmatrix} 0 \\ 0 \\ H_{DC} + H_{AC}\cos\omega_0 t \end{pmatrix}. \tag{4}$$

Including the above selection field and homogeneous stimulation field with an additional offset field,

$$H(x, y, z, t) = H_G(x, y, z) + H_{ST}(t) = \begin{pmatrix} Gx \\ Gy \\ 0 \end{pmatrix} + \begin{pmatrix} 0 \\ 0 \\ H_{DC} + H_{AC}\cos\omega_0 t \end{pmatrix}, \tag{5}$$

which is

$$H(x, y, z, t) = \begin{pmatrix} Gx \\ Gy \\ H_{DC} + H_{AC}\cos\omega_0 t \end{pmatrix}. \tag{6}$$

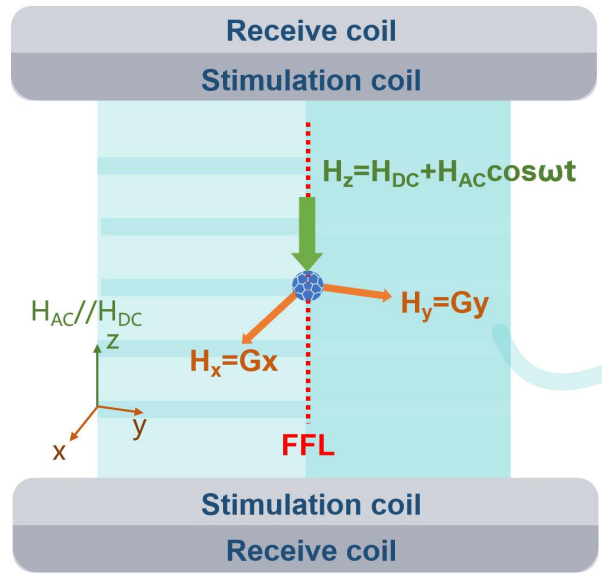

Figure 5. Schematic diagram of MNP stimulation (excitation and depletion), FFL selection, and STED signal detection in STED-MPI system.

*STED Signal*

Using a pair of detection coils in the z direction, the STED signal can be expressed as

$$s_z(t) = -\frac{d}{dt} M_z(t) = -\sum_{k=1}^{3} \frac{\partial M_z}{\partial H_k} \frac{dH_k}{dt}. \tag{7}$$

in which $H_1 = H_x = Gx$, $H_2 = H_y = Gy$, and $H_3 = H_z = H_{DC} + H_{AC}\cos\omega_0 t$. The time derivative of the $H_x$ and $H_y$ component is zero. Because the time derivative of the $H_z$ component is different from zero, the STED signal can be written as

$$s_z(t) = -\frac{\partial M_z}{\partial H_z} \frac{dH_z}{dt}. \tag{8}$$

which is proportional to the time-variation of the z component of the magnetization.

For Langevin particle, the magnetization follows the external field as:

$$M(\xi) = M_0\left(\coth\xi - \frac{1}{\xi}\right), \tag{9}$$

where $M_0$ is the saturation magnetization and

$$\xi = \beta|H| = \frac{\mu_0 m|H|}{k_B T^P}. \tag{10}$$

The field amplitude of the magnetic field $|H|$ can be calculated as

$$|H| = \sqrt{H_x^2 + H_y^2 + H_z^2}, \tag{11}$$

The magnetization $M_z$ along the z direction can be expressed as

$$M_z = M(\xi)\frac{H_z}{|H|} = M_0\left(\coth\xi - \frac{1}{\xi}\right)\frac{H_z}{|H|}. \tag{12}$$

The differential of the magnetization $M_z$ for the z components is:

$$\frac{\partial M_z}{\partial H_z} = \frac{\partial M(\xi)}{\partial H_z}\frac{H_z}{|H|} + M(\xi)\frac{\partial}{\partial H_z}\left(\frac{H_z}{|H|}\right) = M_0\left[\beta\left(\frac{1}{\xi^2} - \frac{1}{\sinh^2\xi}\right)\frac{\partial |H|}{\partial H_z}\frac{H_z}{|H|} + \left(\coth\xi - \frac{1}{\xi}\right)\left(\frac{1}{|H|} - \frac{H_z}{|H|^2}\frac{\partial |H|}{\partial H_z}\right)\right].$$

Based on $\frac{\partial |H|}{\partial H_z} = \frac{H_z}{|H|}$, $\frac{\partial M_z}{\partial H_z}$ can be summarized as

$$\frac{\partial M_z}{\partial H_z} = M_0\left[\beta\left(\frac{1}{\xi^2} - \frac{1}{\sinh^2\xi}\right)\frac{H_z^2}{|H|^2} + \left(\coth\xi - \frac{1}{\xi}\right)\frac{|H|^2 - H_z^2}{|H|^3}\right]. \tag{13}$$

The STED signal is

$$s_z(t) = \omega_0 M_0 H_{AC}\left[\beta\left(\frac{1}{\xi^2} - \frac{1}{\sinh^2\xi}\right)\frac{H_z^2}{|H|^2} + \left(\coth\xi - \frac{1}{\xi}\right)\frac{|H|^2 - H_z^2}{|H|^3}\right]\sin\omega_0 t. \tag{14}$$

A high-pass filter can be used to remove the base frequency component. The unfiltered STED signal amplitude may display a donut-shaped PSF when $H_{DC}$ is greater than $H_{AC}$, however, $H_{DC}$ needs to be much greater than $H_{DC}$ to form a donut-shaped PSF for filtered signal amplitude (Figure 6).

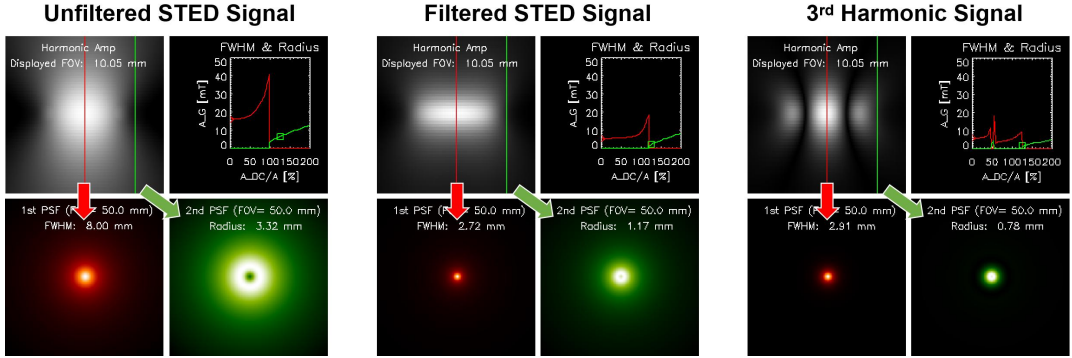

Figure 6. STED signal amplitude and 3rd Harmonic of Langevin particles without relaxation. Gaussian PSF is generated for a small offset magnetic field. Donut PSF can be formed when the offset magnetic field is greater than the excitation field amplitude.

## *STED Signal Frequency Components*

The stimulated emission signal in z direction can be expressed by a Fourier series,

$$s_z(t) = \sum_{n=-\infty}^{\infty} S_{n,z} e^{in\omega_0 t}. \qquad (15)$$

The stimulated emission signal frequency components are defined as

$$S_{n,z} = \frac{1}{T}\int_{-T/2}^{T/2} s_z(t) e^{-in\omega_0 t} dt = -\frac{1}{T}\int_{-T/2}^{T/2} \frac{\partial M_z}{\partial H_z}\frac{dH_z}{dt} e^{-in\omega_0 t} dt. \qquad (16)$$

which is proportional to the time-variation of the $z$ component of the magnetization (Fig. 7).

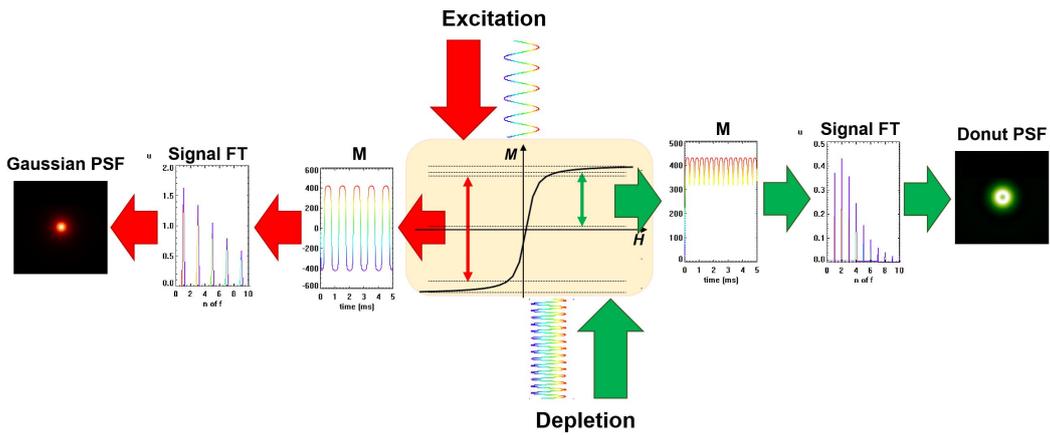

Figure 7. Magnetic particles at the center of an FFL can be stimulated by a dynamic magnetic field with or without a large offset magnetic field. The stimulated emission signal frequency components may exhibit a Gaussian or donut-shaped focal spot on the plane perpendicular to the FFL.

The stimulated emission signal frequency component can be expressed as an integral versus the stimulation magnetic field $H_z$,

$$S_{n,z} = -\frac{1}{T}\int_{H_z(-T/2)}^{H_z(T/2)} \frac{\partial M_z}{\partial H_z} e^{-in\omega_0 t(H_z)} dH_z. \qquad (17)$$

Based on the cosine function of the stimulation field, time can be expressed as

$$t(H_z) = \begin{cases} -\frac{1}{\omega_0} arccos\frac{H_z-H_{DC}}{H_{AC}}, & -\frac{T}{2} \le t \le 0 \\ \frac{1}{\omega_0} arccos\frac{H_z-H_{DC}}{H_{AC}} & 0 \le t \le \frac{T}{2} \end{cases}. \qquad (18)$$

The two solutions can be taken into account excluding the effect of the relaxation time on the magnetization and stimulated emission signal, thus the integral can be split into

$$S_{n,z} = -\frac{1}{T}\left\{\int_{H_{DC}-H_{AC}}^{H_{DC}+H_{AC}} \frac{\partial M_z}{\partial H_z} e^{inarccos\frac{H_z-H_{DC}}{H_{AC}}} dH_z + \int_{H_{DC}+H_{AC}}^{H_{DC}-H_{AC}} \frac{\partial M_z}{\partial H_z} e^{-inarccos\frac{H_z-H_{DC}}{H_{AC}}} dH_z\right\},$$

which can be merged as

$$S_{n,z} = -\frac{1}{T}\int_{H_{DC}-H_{AC}}^{H_{DC}+H_{AC}} \frac{\partial M_z}{\partial H_z}\left(e^{inarccos\frac{H_z-H_{DC}}{H_{AC}}} - e^{-inarccos\frac{H_z-H_{DC}}{H_{AC}}}\right) dH_z. \qquad (19)$$

The real part is zero due to the neglect of the relaxation effects. The stimulated emission signal frequency components only include the imaginary part

$$S_{n,z} = \frac{-2i}{T}\int_{H_{DC}-H_{AC}}^{H_{DC}+H_{AC}} \frac{\partial M_z}{\partial H_z} \sin\left(narccos\frac{H_z-H_{DC}}{H_{AC}}\right) dH_z = \frac{-2i}{T}\int_{H_{DC}-H_{AC}}^{H_{DC}+H_{AC}} \frac{\partial M_z}{\partial H_z} \sin\left(narccos\frac{H_{DC}-H_z}{H_{AC}}\right) dH_z. \qquad (20)$$

Functions of the type $\sin\left(narccos\frac{H_{DC}-H_z}{H_{AC}}\right)$ is similar to Chebyshev polynomials of the second kind:

$$\sin\left(narccos\frac{H_{DC}-H_z}{H_{AC}}\right) = U_{n-1}\left(\frac{H_{DC}-H_z}{H_{AC}}\right)\sin\left(arccos\frac{H_{DC}-H_z}{H_{AC}}\right) = U_{n-1}\left(\frac{H_{DC}-H_z}{H_{AC}}\right)\sqrt{1-\left(\frac{H_{DC}-H_z}{H_{AC}}\right)^2}. \qquad (21)$$

The stimulated emission signal frequency component in $z$ direction ca be written as

$$S_{n,z} = \frac{-2i}{T}\int_{H_{DC}-H_{AC}}^{H_{DC}+H_{AC}} \frac{\partial M_z}{\partial H_z} U_{n-1}\left(\frac{H_{DC}-H_z}{H_{AC}}\right)\sqrt{1-\left(\frac{H_{DC}-H_z}{H_{AC}}\right)^2} dH_z. \qquad (22)$$

Based on the convolution formula $f(H_{DC}) * g(H_{DC}) = \int f(H_z) \cdot g(H_{DC} - H_z) dH_z$, the stimulated emission signal frequency components can be written as a convolution involving the z-direction magnetic field $H_z$ only, i.e.

$$S_{n,z}(H_{DC}) = \frac{-2i}{T}\frac{\partial M_z}{\partial H_z}(H_{DC}) * \left(U_{n-1}\left(\frac{H_{DC}}{H_{AC}}\right)\sqrt{1-\left(\frac{H_{DC}}{H_{AC}}\right)^2}\right). \qquad (23)$$

The stimulated emission signal frequency component is the derivative of the magnetization convolved with a set of Chebyshev polynomials. The convolution kernel part $U_{n-1}\left(\frac{H_{DC}}{H_{AC}}\right)\sqrt{1-\left(\frac{H_{DC}}{H_{AC}}\right)^2}$ is zero when $H_{DC} > H_{AC}$.

### *STED Signal Frequency Components of the Langevin Particles*

For the Langevin particle, the spatial dependency of the stimulated emission signal picked up in z-direction ($S_{n,z}$) becomes

$$S_{n,z} = \frac{-2iM_0}{T}\left[\beta\left(\frac{1}{\xi^2} - \frac{1}{\sinh^2\xi}\right)\frac{H_{DC}^2}{|H|^2} + \left(\coth\xi - \frac{1}{\xi}\right)\frac{|H|^2 - H_{DC}^2}{|H|^3}\right] * \left(U_{n-1}\left(\frac{H_{DC}}{H_{AC}}\right)\sqrt{1-\left(\frac{H_{DC}}{H_{AC}}\right)^2}\right), \quad (24)$$

The field amplitude of the magnetic field for $H_z = H_{DC}$ is

$$|H| = \sqrt{(Gr)^2 + H_{DC}^2} = G\sqrt{r^2 + (H_{DC}/G)^2}, \quad (25)$$

where $r = \sqrt{x^2 + y^2}$ is the distance of the point to the original point in x-y plane. The spatial dependency of the stimulated emission signal picked up in z-direction ($S_{n,z}$) becomes

$$S_{n,z}(r, H_{DC}) = \frac{-2iM_0}{T}\left[\frac{\mu_0 m}{k_B T^p}\left(\frac{1}{\xi^2} - \frac{1}{\sinh^2\xi}\right)\frac{(H_{DC}/G)^2}{[r^2+(H_{DC}/G)^2]} + \left(\coth\xi - \frac{1}{\xi}\right)\frac{r^2}{G[r^2+(H_{DC}/G)^2]^{3/2}}\right] * \left[U_{n-1}\left(\frac{H_{DC}}{H_{AC}}\right)\sqrt{1-\left(\frac{H_{DC}}{H_{AC}}\right)^2}\right] \quad (26)$$

in which the convolution involves the $H_{DC}$ variable only (Fig. 8).

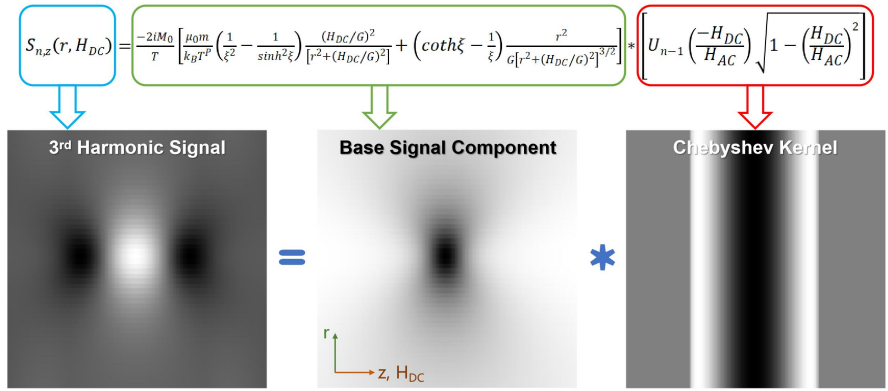

Figure 8. The 3rd Harmonic signal component for Langevin particles under stimulation with an offset can be calculated by the convolution of the base signal and Chebyshev kernel.

A regular Gaussian focal spot can be formed when there is no offset magnetic field $H_{DC} = 0$. By increasing the offset magnetic field $H_{DC}$ (with $H_{DC} > H_{AC}$), the stimulated emission signal is depleted as the center of the donut focal spot (FFL center). The stimulated emission signal increases and then decreases towards the orthogonal gradient $H_G$ direction, forming a donut focal spot. in which the convolution involves the $H_{DC}$ variable only (Fig. 9).

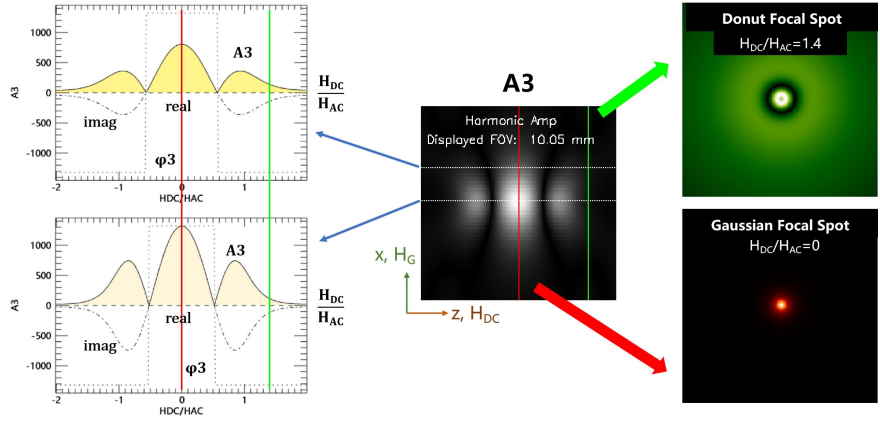

Figure 9. The 3rd Harmonic STED signal component exhibits different focal spot shapes depending on the DC field amplitude, e.g. a Gaussian bright focal spot when $H_{DC} = 0$ and a donut focal spot with depletion in the center when $H_{DC} > H_{AC}$.

## *STED Signal of Debye Particles*

The relaxation effects of the Langevin particles may induce a magnetization and signal lag (Fig. 10) during the excitation[35]. The STED signal of Debye particles induced in the receiver coils can be expressed as follows[36]:

$$s_{z\_Debye}(t) = -\frac{d}{dt}(M_z(t) * R(t)), \tag{27}$$

where $R(t)$ is the Debye relaxation kernel of magnetic particles[37] and defined as follows:

$$R(t) = \frac{1}{\tau} exp\left(\frac{-t}{\tau}\right) u(t). \tag{28}$$

$u(t)$ is the Heaviside step function stepping at $t = 0$. The relaxation time $\tau$ can be approximate as the characteristic zero-field Brownian relaxation time[31]

$$\tau \approx \frac{3\eta V_h}{k_B T^p}, \tag{29}$$

where $\eta$ represents the fluid viscosity and $V_h$ represents the hydrodynamic volume of the MNPs[38].

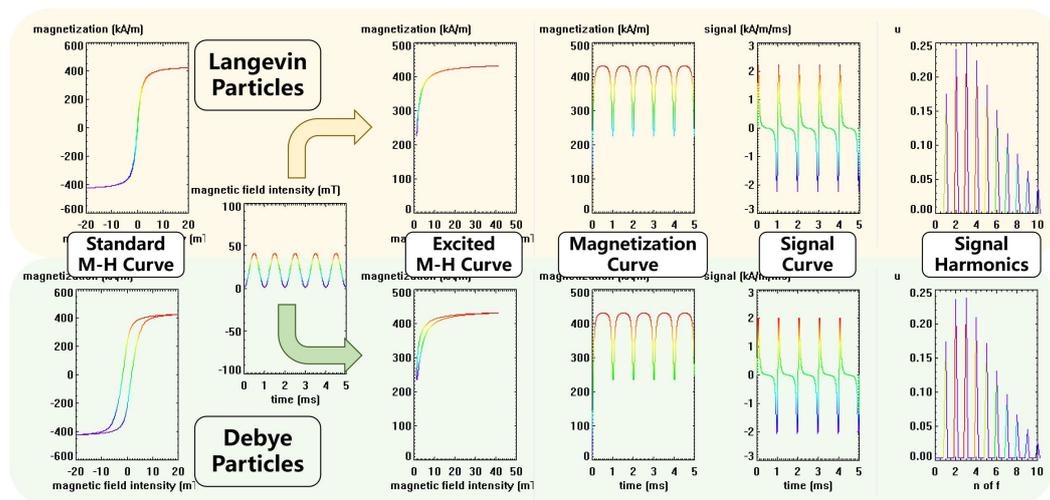

Figure 10. STED signal generated from Langevin and Debye particles.

Similarly, we can calculate the original and filtered STED signals, as well as the signal frequency components of the magnetic particles with relaxation. Depending on the offset magnetic field, either Gaussian or Donut focal spots may be generated on the orthogonal gradient field plane (Figure 11).

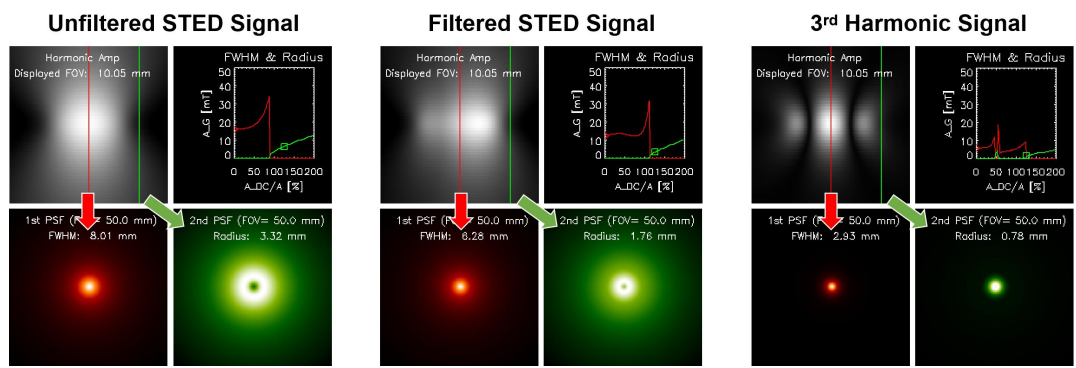

Figure 11. STED signal amplitudes and 3rd Harmonic signal of Langevin particles with Debye relaxation. The donut radius of the filtered signal becomes greater due to the Debye relaxation. The donut radius of the harmonic signals is not affected by the Debye relaxation.

## STED-MPI Signal Simulation

The STED signal simulation process includes 5 parts: parameter setting, time domain analysis, frequency domain analysis, STED PSF formation, and image reconstruction (Fig. 12). The STED simulation was performed using in-house developed software based on the Interactive Data Language (IDL, Exelis Visual Information Solutions, Boulder, CO, USA).

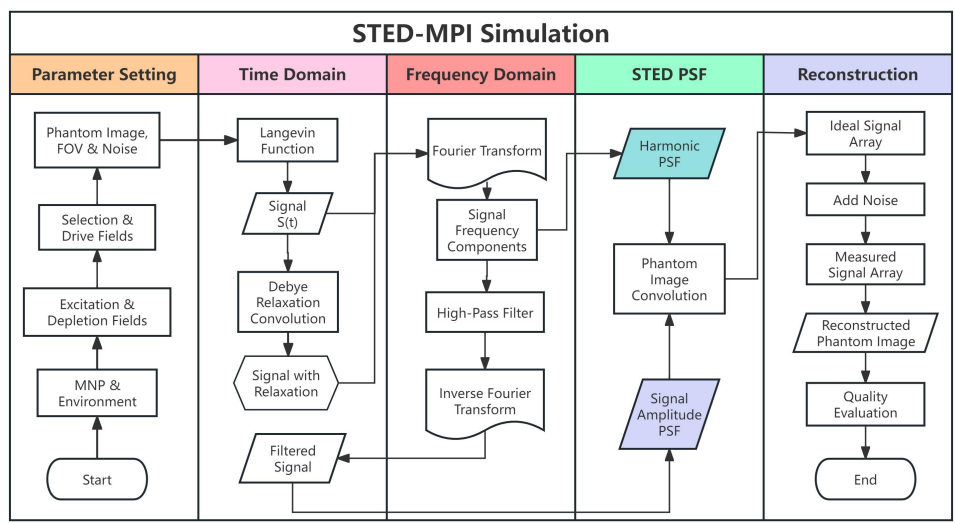

Figure 12. STED-MPI simulation process. Donut PSF can be generated by simulating phantom scans and image reconstruction processes for high-resolution MPI.

The excitation magnetic field in Eq. 2 was set between 1 to 100 mT with a frequency ranging within 100 kHz. The sampling rate is 1 MHz. The offset magnetic field in Eq. 3 was set between -200% and 200% of the excitation magnetic field amplitude. Two offset magnetic fields were used to generate Gaussian and donut focal spots respectively. The selection gradient field in Eq. 1 was set ranging from 0.1-10 T/m. After the Fourier transformation of the signal, the amplitude, phase, real part, and imaginary part of the signal frequency components (1$^{st}$ to 31$^{st}$ Harmonics) are used to generate PSFs (Fig. 13).

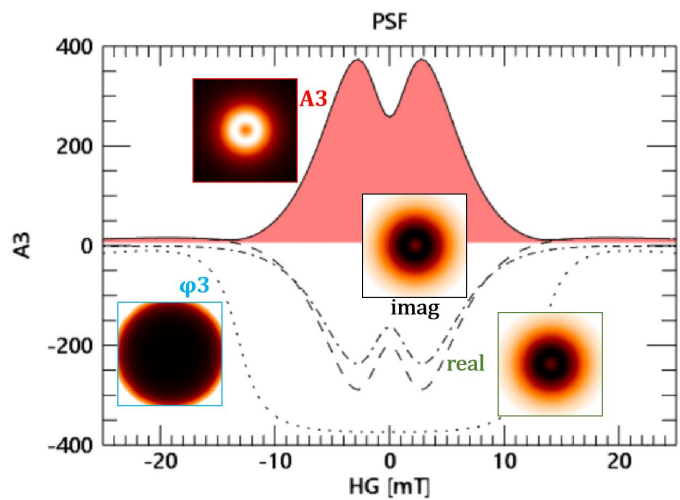

Figure 13. The donut focal spot may be formed by the amplitude of the 3rd harmonic frequency component (A3) when the offset magnetic field is 150% of the excitation magnetic field.

We simulated a set of magnetic nanoparticles with particle sizes ranging from 10–100 nm. Langevin function was used to estimate the adiabatic signal, from which the non-adiabatic signal was calculated using the Debye relaxation model. The viscosity was set ranging from 1 to 50 mPa·s, and the temperature was 0-100 ℃. Brownian relaxation time can be calculated from these parameters by using Equation 30. The signal frequency components from Chebyeshev theory, Langevin particles, and Debye particles are compared in this study (Fig. 14).

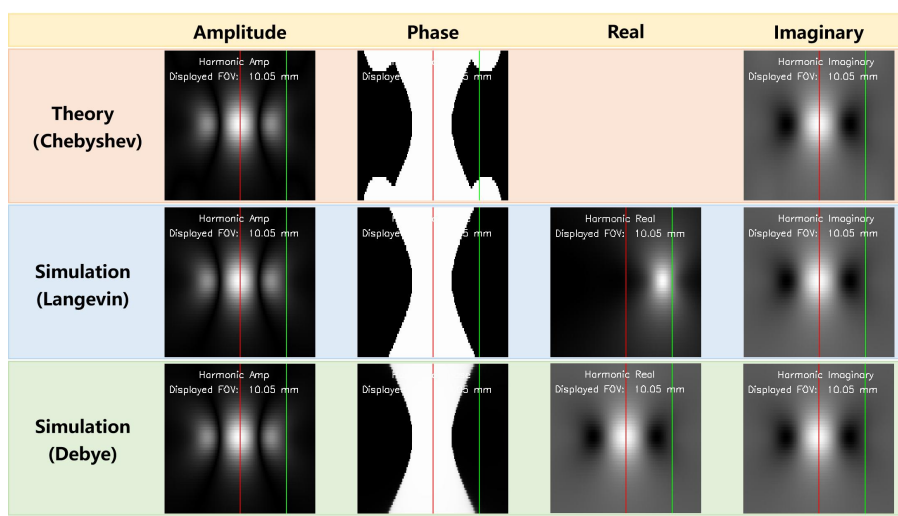

Figure 14. Signal frequency component A3 from theory and simulations.

In the simulation, as the offset magnetic field $H_{DC}$ increases, the magnetic particle signal distribution map in the orthogonal gradient selection field $H_G$ direction is displayed as different magnetic particle focal spots. The orthogonal gradient selection field $H_G$ in the x and y directions is added to collect the above magnetic particle signal distribution maps. When the offset magnetic field $H_{DC}$ increases, in the x-y plane applied by the gradient magnetic field $H_G$, the magnetic particle focal spot changes from a large bright spot to a bright ring focal spot, a dark ring focal spot, a small bright spot, a small bright ring focal spot, and finally the bright ring focal spot keeps increasing (Fig. 15).

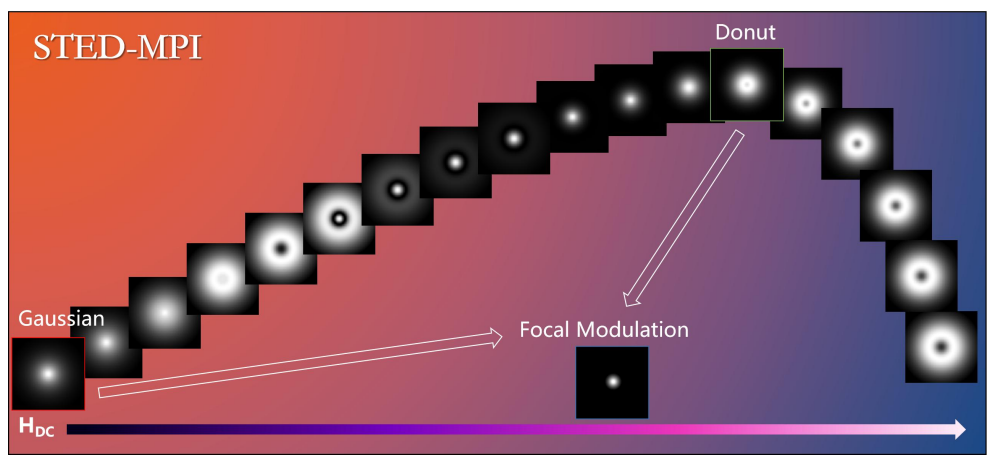

Figure 15. Focal spots with different shapes and sizes may be formed by increasing the amplitude of the offset magnetic field.

By varying the excitation and offset magnetic fields, either Gaussian PSFs (Fig. 16) or donut PSFs (Fig. 17) can be formed along the orthogonal gradient selection field directions.

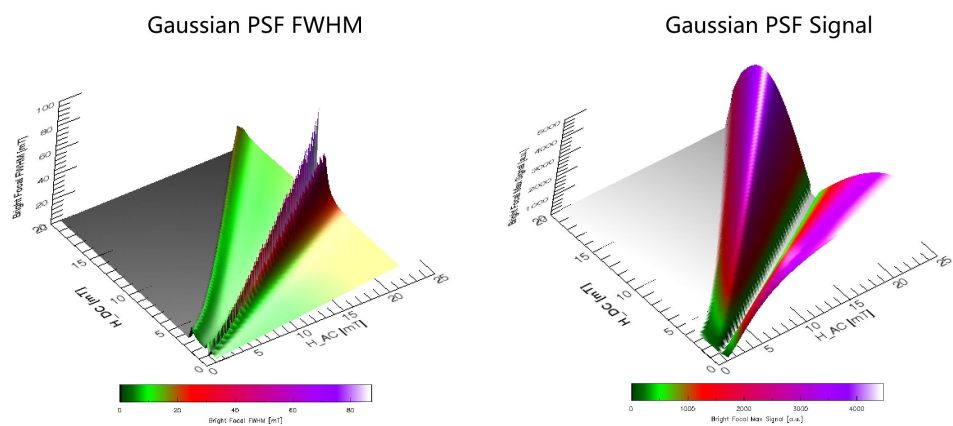

Figure 16. The FWHM and signal relative signal of Gaussian PSFs at different excitation and offset magnetic fields.

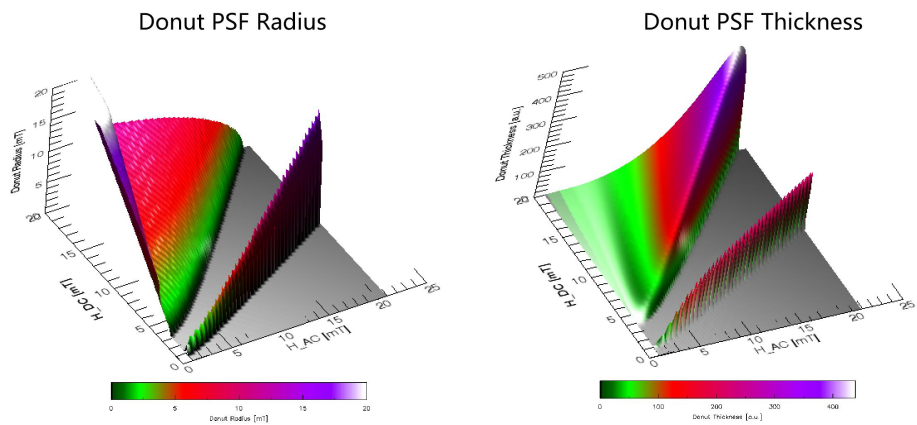

Figure 17. The radius and thickness of donut PSFs at different excitation and offset magnetic fields.

## STED-MPI Scanner Simulation

The simulation scanner based on STED-MPI mainly includes the following five modules (Fig. 18): 1) Excitation magnetic field module. A uniform magnetic field excitation unit provides a constant uniform alternating magnetic field in the z-direction. A constant uniform alternating magnetic field is an alternating magnetic field with constant uniform strength in its magnetic field direction and the magnetic field in the z-direction. The uniform magnetic field excitation unit is provided with a power interface, to which the control module provides a current and controls the power signal. The focal spot-modulated magnetic field excitation coil pairs in the x-direction and y-direction are used to provide the gradient magnetic field in the respective direction. 2) Control module: The control module is mainly used for current control, including the waveform generator and its corresponding front-end controller. The waveform generator can raise the city voltage to a certain value, change the boosted alternating current into DC through rectification, and obtain alternating current at a certain frequency through the frequency converter, such as the frequency range of 2-50 kHz; The front-end controller drives the scanning sequence and distributes current to each coil under the high voltage control of the frequency conversion output. In addition, the current size applied to the respective coil can be fed back before the pre-drive through a feedback loop to form a closed-loop control. 3) Signal receiving module: The change of magnetic flux caused by the magnetization response of magnetic nanoparticles is received to generate the corresponding induced voltage signal. 4) Signal processing module: The voltage signal detected from the receiving coil pair is processed. The signal processing module may include an analog processing sub-module and a digital processing sub-module. 5) Image reconstruction module: process and reconstruct the signal to obtain the spatial distribution image of the magnetic particle concentration.

STED-MPI can be implemented on available FFL-based MPI scanner systems (Fig. 18). One example scanner includes a vertical FFL which generates permanent NdFeB magnets and shifted by shift coils for 2D and three-dimensional (3D) scans[39]. Another example scanner is the open-sided MPI system with electronically rotating FFL[40]. The last, but not least scanner example is the open-bore narrow-band scanner based on a double-layer linear scanning structure, in which a fully mirrored coil pack is added to the original gradient and shift coils to enhance the linearity of FFL[13]. STED-MPI can be achieved by applying excitation and depletion fields along with the shifting or rotating FFL. Receiver coils can collect either Gaussian or donut-shaped PSF signals from magnetic nanoparticles for high-resolution imaging.

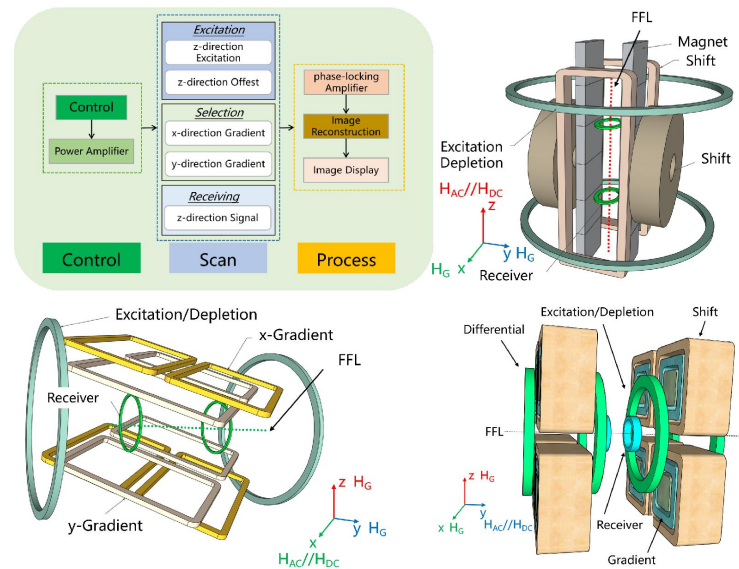

Figure 18. STED-MPI system component diagram and example scanners.

We can improve the image resolution by subtracting the image of the donut focal spot from the image of the smallest Gaussian focal spot. Fig. 19 shows two separated magnetic particle points corresponding to two focal spots. The red curves in the images of the second column are the signal distribution map given by the dotted line of the middle figure in the first row, and the blue curve is the real position distribution of the magnetic particle points. It can be seen that the focal-modulated image has smaller focal spots and a more accurate position of the magnetic particle distribution.

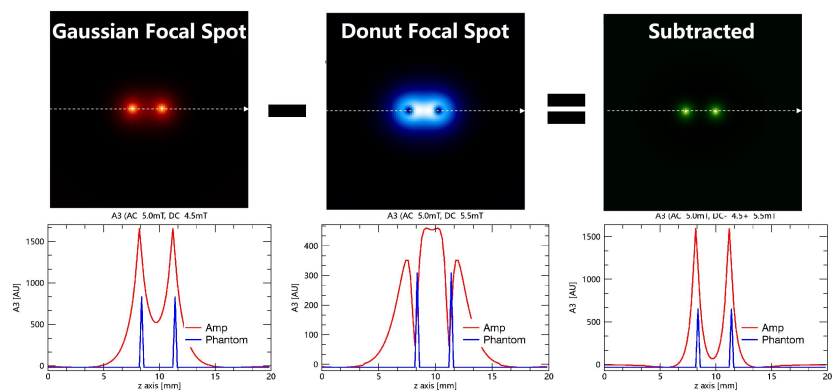

Figure 19. High-resolution images can be derived by modulating Gaussian and donut focal spot images in STED-MPI.

Phantom images (Fig. 20) include multiple point sources, 2D bar pattern phantom, bars with different concentrations, ACR mammography phantom, Shepp-Logan phantom, brain vessel projection maps from TOF-MRA images[41], vascular phantom, and neoplasm phantom[31]. Measured signal images are generated based on the convolution of a phantom image with the Gaussian and donut PSFs, after which noise is added to mimic the electronic signal collection process.

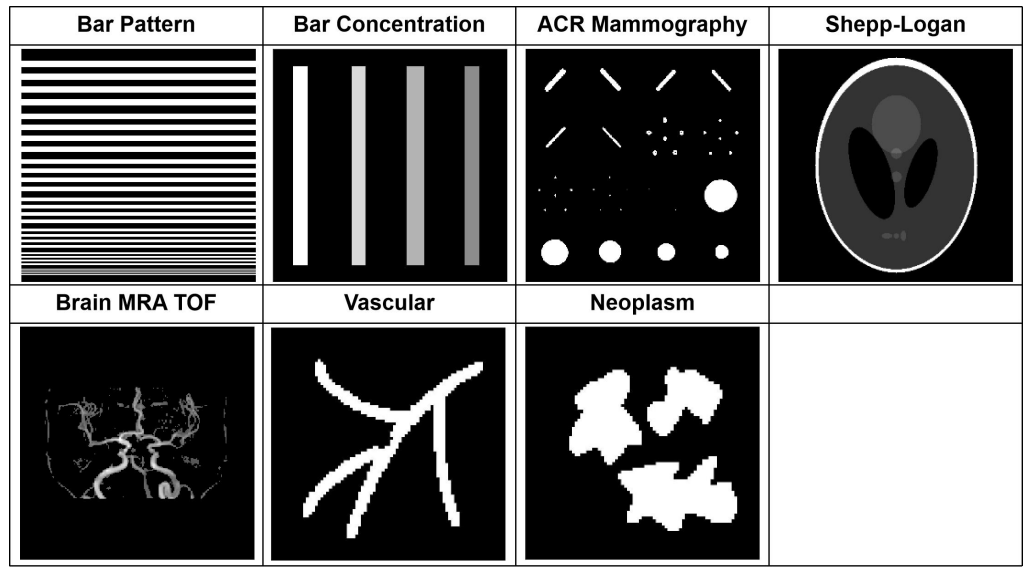

Figure 20. Phantoms used in this study for STED-MPI examination.

A deconvolution technique based on the Fourier transform was used to generate higher-resolution images (Fig. 21). The first step is to obtain the k-space image of the Gaussian- and donut-based phantom signal maps through Fourier transform. The second step is performed through the Fourier transform to generate the two-dimensional k-space images of the Gaussian and donut focal spots. The third step is to divide the k-space image of the phantom signal by the k-space image of the focal spots. The fourth step is performed through the inverse Fourier transform to reconstruct the phantom images from the Gaussian and donut focal spots. The last step is performed by subtracting the reconstructed images. In Fig. 21, the red curve is the k-space distribution curve of the focal spot, the blue curve is the k-space distribution curve of the phantom signal image, and the green curve is the k-space distribution curve by dividing the blue curve by the red curve, which is the crucial step of the deconvolution-based reconstruction technique. A threshold value was used to filter the background noise pixels in the reconstructed STED-MPI images.

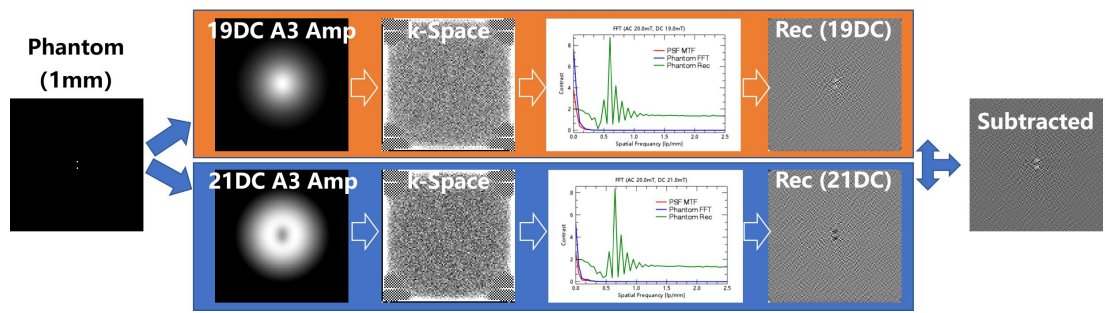

Figure 21. High-resolution images can be reconstructed by Fourier-transform-based deconvolution in STED-MPI.

In the simulation, the Gaussian focal spot and the donut focal spot may be subtracted to generate a smaller PSF (Fig. 22). A dog (difference of Gaussian) factor is used to further shrink the PSF during the subtraction of two focal spots, which is the so-called Gaussian transformation:

$$PSF_{sub} = PSF_1 - f_{dog} \cdot PSF_2 \tag{30}$$

In this study, a 0% DC offset was set as the Gaussian PSF. A 105% DC offset was selected as a donut PSF and a dog factor of 1.9 was chosen for a smaller PSF generation and deconvolution-based image reconstruction.

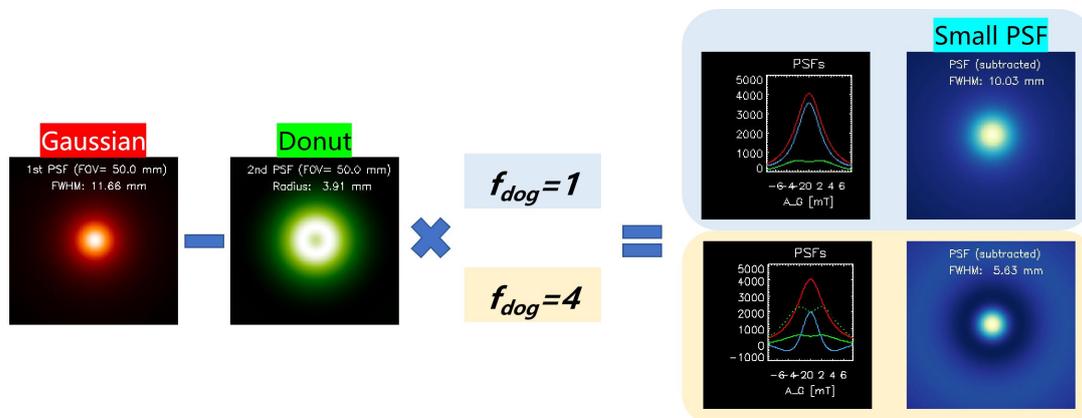

Figure 22. A dog factor can be used to generate a much smaller PSF than the bright Gaussian PSF during the focal modulation process.

Reconstructed image quality was evaluated by using root mean square error (RMSE), peak signal-to-noise ratio (pSNR), and structure similarity (SSIM), as shown in Table 2.

| Table 2. Reconstructed image quality assessment in STED-MPI. | | | | | | | | | |
|---|---|---|---|---|---|---|---|---|---|
| Phantom | RMSE | | | pSNR | | | SSIM | | |
|  | Gaussian | Donut | STED | Gaussian | Donut | STED | Gaussian | Donut | STED |
| Bar Pattern | 10.22 | 10.26 | 7.78 | 64.36 | 64.30 | 71.18 | 0.033 | 0.019 | 0.192 |
| Bar Concentration | 9.80 | 9.08 | 5.50 | 65.24 | 66.84 | 76.68 | 0.177 | 0.215 | 0.710 |
| ACR Mammo | 7.73 | 7.35 | 2.49 | 69.97 | 71.14 | 92.78 | 0.228 | 0.272 | 0.860 |
| Shepp-Logan | 8.85 | 8.94 | 5.97 | 67.26 | 67.05 | 74.91 | 0.296 | 0.302 | 0.523 |
| Brain MRA TOF | 6.27 | 6.58 | 3.16 | 74.07 | 72.94 | 88.33 | 0.398 | 0.363 | 0.825 |
| Vascular | 7.77 | 8.76 | 4.00 | 69.52 | 67.25 | 82.92 | 0.263 | 0.182 | 0.851 |
| Neoplasm | 9.91 | 9.31 | 4.90 | 67.18 | 66.15 | 79.21 | 0.256 | 0.220 | 0.695 |

## STED Signal Experiment

The large offset magnetic field inducing a donut-shaped focal spot was observed on the MPI scanner (Fig. 23)[43]. The scanner used two offset field amplitudes (8 and 14.2 mT). The gradient field along with the excitation field direction (x-direction) was 1.717 mT/mm. The z-direction gradient field was 3.434 mT/mm. The imaging FOV was 30*30 mm$^2$.

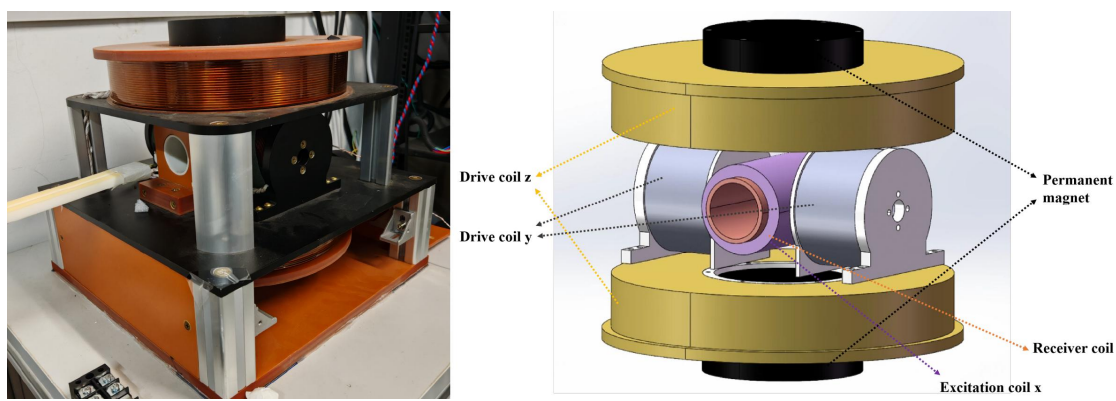

Figure 23. Photo (Left) and design (Right) of MPI scanner for donut focal spot observation.

The donut-shaped focal spot with two peaks can be detected when the offset magnetic field is greater than the excitation magnetic field. When the excitation field is 14.2 mT, the two peaks are clearly shown in Fig. 24.

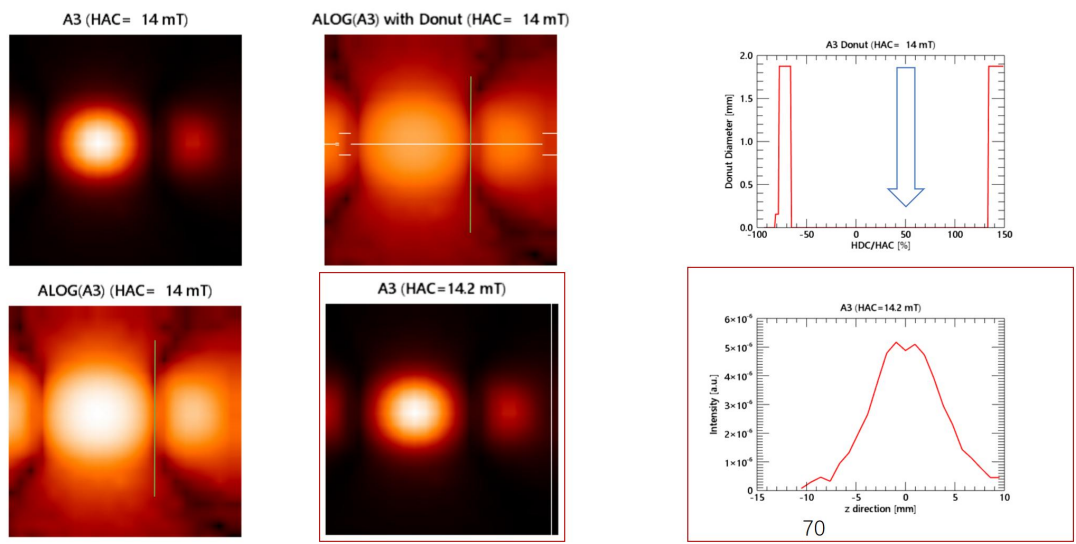

70

Figure 24. The line profile with two peaks along the gradient field direction is shown when the offset field is greater than the excitation field.

## Discussions

In this study, a donut-shaped focal spot is generated by using an excitation field with an offset field, which is similar to a depletion field in STED fluorescence microscopy[44]. However, the donut-shaped focal intensity distribution with zero intensity in the center in STED microscopy is created by a stimulated laser beam via a vortex phase mask. We provide a spatially homogeneous excitation and offset magnetic fields, center signal is depleted in STED-MPI due to the bypassing of the center of Langevin magnetization function. When the offset field is smaller than the excitation field amplitude, the signal is usually shown as a Gaussian shape. The donut-shaped signal cannot be created until the offset field is greater than the excitation field. The donut radius gradually increases when the offset magnetic field increases. The Gaussian and Donut focal spots may be combined to improve the resolution of MPI.

In the simulation, we introduced noise into the convoluted signal maps by applying normally distributed noise to the entire map, which may be improved by introducing noise into every signal acquisition part.

MNP relaxation does not affect on the donut radius of the signal frequency components. We used the Debye relaxation model and Brownian relaxation time to mimic the magnetization lag in excitation. Both Néel and Brownian relaxation times[45] may be included to improve an accurate analysis of the donut shape transformation in STED-MPI.

The donut-shaped focal spot usually has a small SNR due to the large offset magnetic field. A higher excitation frequency may be used to improve the SNR because the noise in the receiver electronics is dominated by a $1/f$ factor. However, the body energy deposition or specific absorption rate may pose limitations to the use of higher excitation frequencies[12].

Donut-shaped focal spot exists on both signal amplitude and signal frequency components. The former provides a higher SNR but a larger donut, the latter provides a smaller donut with a lower SNR. It is crucial to combine them for a better image resolution in STED-MPI.

## Conclusion

This study proposes an innovative idea of high-resolution MPI based on a donut PSF and STED microscopic imaging principle. The STED signal in MPI can be generated by adding a large offset magnetic field parallel to the FFL, which may form a donut-shaped focal spot or a regular Gaussian focal spot depending on the offset field strength. Focal spot modulation techniques and deconvolution algorithms are developed to improve the image resolution in MPI.

28